\documentclass[5p, twoclumn]{elsarticle}

\usepackage{amsfonts,amssymb,amsmath,times,enumerate}
\usepackage{subfigure}
\usepackage{graphicx,color}

\begin{document}

\title{Geometric detection of coupling directions by means of inter-system recurrence networks}

\author[pik,hub]{Jan H. Feldhoff}
\author[pik]{Reik V. Donner\corref{cor1}}
\ead{reik.donner@pik-potsdam.de}
\author[pik,hub]{Jonathan F. Donges}
\author[pik]{Norbert Marwan}
\author[pik,hub,uab]{J\"urgen Kurths}
\cortext[cor1]{Corresponding author. Phone/fax: +49-331-288-2064/2640}
\address[pik]{Potsdam Institute for Climate Impact Research, P.O. Box 60\,12\,03, 14412 Potsdam, Germany}
\address[hub]{Department of Physics, Humboldt University, Newtonstr.~15, 12489 Berlin, Germany}
\address[uab]{Institute for Complex Systems and Mathematical Biology, University of Aberdeen, Aberdeen AB243UE, United Kingdom}

\date{\today}

\begin{abstract}
We introduce a geometric method for identifying the coupling direction between {two} dynamical systems based on a {bivariate} extension of recurrence network analysis. Global characteristics of the resulting inter-system recurrence networks provide a correct discrimination for weakly coupled R\"ossler oscillators not yet displaying generalised synchronisation. Investigating two real-world palaeoclimate time series representing the variability of the Asian monsoon over the last 10,000 years, we observe {indications for} a considerable influence of the Indian summer monsoon on climate in Eastern China rather than vice versa. {The proposed approach can be directly extended to studying $K>2$ coupled subsystems.} \\


\end{abstract}

\maketitle

\section{Introduction}

Uncovering causal interdependencies from observational data is one of the great challenges of nonlinear time series analysis \cite{Palus_pre_2007}. While detecting statistical interrelationships between interacting systems is comparably simple and can be achieved by many linear as well as nonlinear methods such as cross-correlation, mutual information \cite{Frenzel_prl_2007}, or synchronisation analysis \cite{Pikovsky_Kurths_synchr}, the correct identification of coupling direction is still faced with numerous practical problems \cite{Palus_pre_2007}. Existing methods for this purpose include linear Granger causality \cite{Granger1969} as well as nonlinear extensions thereof \cite{Ancona2004,Chen2004}, information-theoretic quantities like transfer entropy \cite{schreiber_prl2000,staniek_prl2008} or conditional mutual information \cite{vejmelka_pre_2008}, and state-space based characteristics such as conditional probabilities of recurrences \cite{Romano_pre_2007,zou2011}, to mention only a few examples. In general, all methods proposed so far make explicit use of \textit{dynamical} characteristics, and it is still a matter of scientific debate which method performs best under specific conditions. In turn, the potential \textit{geometric} consequences of directed couplings in the shared phase space of interacting dynamical systems have not been explicitly studied so far. This work presents a first attempt to a corresponding structural characterisation.

In the last years, several approaches for analysing time series from dynamical systems by means of graph-theoretical concepts have emerged~\cite{Zhang2006,Xu2008,Lacasa2008,Yang2008,Marwan2009} (see \cite{Donner2010NJP,Donner2011IJBC} for a corresponding review). Among these different approaches, networks based on the mutual proximity of state vectors in phase space have become increasingly popular, since their structural properties are closely related to the geometry of the underlying attractor and, hence, the resulting dynamics.

One particularly important class of proximity networks in phase space are recurrence networks~\cite{Marwan2009,Donner2010NJP,Donner2011IJBC}, which are especially useful for characterising the structural properties of low-dimensional systems. Recall that the property of recurrence (in the sense of Poincar\'e) corresponds to geometrical closeness in phase space without the necessity of trivial temporal correlations. This closeness can be characterised in different ways, including neighbourhoods of individual states with fixed probability mass ($k$-nearest neighbour approach) or with a fixed phase space volume (determined by a maximum spatial distance $\varepsilon$ of neighbouring states). In the latter case, the recurrence properties obtained from a particular realisation (i.e., a -- possibly multivariate -- time series) $\{x_i\}_{i=1}^N$ representing the relevant degrees of freedom of a complex system $X$ are encoded in the binary recurrence matrix
\begin{equation}
R_{ij}(\varepsilon)=R^X(x_i,x_j|\varepsilon)=\Theta(\varepsilon-d(x_i,x_j)),
\label{def:epsrec}
\end{equation}
\noindent
where $d(\cdot,\cdot)$ measures some distance (e.g.,~according to the Euclidean or maximum norm) in phase space, and $\Theta(\cdot)$ denotes the Heaviside function. The visualisation of this symmetric matrix is known as the \textit{recurrence plot}~\cite{Eckmann1987}, and the emergence of line structures in such recurrence plots has been intensively utilised for characterising the dynamical properties of the underlying time series by estimates of dynamical invariants and novel measures of complexity (\textit{recurrence quantification analysis, RQA})~\cite{marwan2007}.

Recurrence network analysis reinterprets the recurrence structure of a time series as the connectivity pattern of an associated complex network represented by an undirected simple graph \cite{Marwan2009}. Specifically, given the definition in Eq.~(\ref{def:epsrec}) based on $\varepsilon$-recurrences, we can formally write
\begin{equation}
A_{ij}(\varepsilon)=R_{ij}(\varepsilon)-\delta_{ij}
\end{equation}
\noindent
(where $\delta_{ij}$ is Kronecker's delta) to obtain the adjacency matrix of the corresponding $\varepsilon$-recurrence network (we will abbreviate this term by RN in the following). The properties of such networks have been widely studied elsewhere~\cite{Donner2010NJP,Donner2011IJBC,Donner2011EPJB,Zou2010,Donges2011PRE,Zou2011Chaos,Strozzi2011}, and their practical use as an exploratory tool of time series analysis has been demonstrated~\cite{Marwan2009,Donges2011PNAS,Donges2011NPG}. In particular, global properties of RNs have been shown to trace dynamical transitions in non-stationary time series~{\cite{Marwan2009,Donner2011IJBC,Donges2011PNAS,Donges2011NPG}}. A particularly relevant, yet straightforward characteristic is the edge density 
\begin{equation}
\rho=\frac{1}{N(N-1)}\sum_{i,j} A_{ij},
\end{equation}
\noindent
which equals (for a RN) the recurrence rate $RR$ of the underlying time series. We emphasise that there is a simple monotonic relationship between $\rho$ and the recurrence threshold $\varepsilon$. In the context of RNs~{\cite{Donges2011PRE,Donner2010PRE}} (but also other spatially embedded graphs such as climate networks~\cite{Donges2009a,Donges2009b}), it has been argued that a suitable resolution of small-scale network features requires the choice of a low edge density, typically $\rho\lesssim 0.05$, confirming earlier suggestions provided 
for recurrence plots \cite{marwan2007,Schinkel2008,marwan2011}.

In previous research, the concept of RNs has only been applied for the analysis of single (uni- or multivariate) time series or for the characterisation of individual dynamical systems, whereas the underlying recurrence plot concept has been complemented by bi- and multivariate extensions \cite{marwan2007}. In this work, we provide {one possible} framework for {bi- and} multivariate RN analysis and discuss its potentials for detecting the coupling direction between two or more complex dynamical systems. Although our considerations are particularly made for RNs, similar conceptual developments should be possible for other types of recurrence networks (e.g., $k$-nearest neighbour networks). 

This paper is organised as follows: In Section 2, we present our methodological approach to a joint treatment of multiple time series in a RN framework. The associated characteristics describing some fundamental structural properties of general coupled networks (sometimes also referred to as interconnected, interdependent, interacting or \textit{network of networks}~\cite{Buldyrev2010} in the literature) are reviewed. Subsequently, we outline how these properties can be used for characterising possible geometric signatures of coupling between dynamical systems. The proposed framework is then illustrated for two {diffusively} coupled R\"ossler systems as a paradigmatic example in Section 3. The effects of different dynamical regimes, the emergence of generalised synchronisation, and the lengths of the available time series on the detectability of the true coupling direction are discussed in some detail. In addition, a real-world example from climatology is presented in Section 4. Finally, the main conclusions from our work are summarised in Section 5.

\section{Inter-system recurrence networks and coupling detection}

\subsection{Cross-recurrence plots}

Let $\{x_i\}_{i=1}^{N_X}$ and $\{y_j\}_{j=1}^{N_Y}$ represent realisations of two dynamical systems $X$ and $Y$ observed at distinct times $t_i$ ($i=1,\dots,N_X$) and $t_j$ ($j=1,\dots,N_Y$), respectively{, which do not need to be the same for both systems}. This is, $x_i=X(t_i)$ and $y_j=Y(t_j)$. If both systems share a common phase space (i.e., $X$ and $Y$ represent the same physical quantities), we can consider a distance between individual observations (state vectors) of both systems as
\begin{equation}
d^{XY}(x_i,y_j)=d^{XY}(X(t_i),Y(t_j))=\|x_i-y_j\|,
\end{equation}
\noindent
where $\|\cdot\|$ is a suitable norm (e.g., Euclidean norm) in phase space. Then, the cross-recurrence matrix is \cite{ Marwan2002PLA} 
\begin{equation}
\begin{split}
CR_{ij}(\varepsilon^{XY})&=CR^{XY}(x_i,y_j|\varepsilon^{XY})\\
&=\Theta(\varepsilon^{XY}-d^{XY}(x_i,y_j)).
\end{split}
\label{def:crp}
\end{equation}
\noindent
Unlike the univariate recurrence matrix $\mathbf{R}$, $\mathbf{CR}$ is in general not symmetric and not even necessarily a square matrix (i.e., $N_X\neq N_Y$ is possible). As in the univariate case, the cross-recurrence rate
\begin{equation}
RR^{XY}=\sum_{i,j} CR^{XY}_{ij}
\end{equation}
\noindent
is an important characteristic of the cross-recurrence matrix $\mathbf{CR}$ that increases montonically with $\varepsilon^{XY}$.

The statistical distribution of line structures in the associated cross-recurrence plot {(CRP)} can be utilised for defining bivariate measures of complexity to study interrelationships between possibly coupled dynamical systems \cite{Marwan2002PLA,marwan2002npg}. However, this point of view typically requires that the sampling of both systems is comparable, since otherwise, line structures are difficult to define.

{From a conceptional perspective, we have to} emphasise {that for the formal applicability of CRPs,} the {two studied} systems must share the same {(reconstructed)} phase space, i.e., they {need to} have the same set of dynamically relevant variables \cite{marwan2007,marwan2011}. {Although there is no formal proof that this prerequisite provides a generally applicable sufficient or even necessary condition, it provides a reasonable working basis that can be considered as being} typically met when comparing observations with the same physical meaning, such as temperature records from two distant observatories or EEG measurements from different electrodes. {In turn}, in many real-world problems, {one is rather interested in studying interrelationships between two physically different quantities, for which} this requirement is not fulfilled. {However, in many applications, the resulting} conceptual problems have {been widely ignored} by rendering the {respective phase spaces} of two systems {quantitatively similar} by means of suitable monotonic transformations of the data (e.g., standardisation to zero mean and unit variance \cite{Marwan2002PLA,marwan2003climdyn}, quantile adjustment, or distinguishing positive and negative interdependencies \cite{Donner2011Extremis,marwan2003climdyn}). An alternative solution has been proposed by using order patterns, which allows us to directly compare time series in terms of their local rank order \cite{groth2005}, {but destroys much of the quantitative information usually contained in recurrence plot-based techniques. Consequently,} results based on the intercomparison of two structurally different variables {by means of CRPs} should {always} be interpreted and discussed carefully.

\subsection{Inter-system recurrence networks}

{In the following}, we {combine} the information on recurrences within the individual dynamical systems {$X$ and $Y$} (i.e., ``intra-system'' recurrences, Eq.~(\ref{def:epsrec})) with that on their respective cross-recurrences (i.e., proximities between pairs of states of {both} systems, Eq.~(\ref{def:crp})) in an \textit{inter-system recurrence matrix}
{
\begin{equation}
\mathbf{IR}(\varepsilon)=\left( \begin{array}{cc} \mathbf{R}^X(\varepsilon) & \mathbf{CR}^{XY}(\varepsilon) \\
\mathbf{CR}^{YX}(\varepsilon) & \mathbf{R}^Y(\varepsilon) \end{array} \right)
\label{isrm}
\end{equation}
with $\varepsilon$ being a common threshold for defining the spatial proximity between the state vectors of each as well as both systems. $\mathbf{IR}$} can also be derived as the {standard} recurrence matrix of the concatenated {observations   $\mathbf{x}=(x_{1},\dots,x_{N_X},y_{1},\dots,y_{N_Y})$}. 
Moreover, {since $\mathbf{CR}^{YX}=[\mathbf{CR}^{XY}]^T$}{, $\mathbf{IR}$ is a symmetric matrix}.  

{We emphasise that the} adjacency matrix of {two} coupled networks (see~\cite{Donges2011EPJB}) can be written in complete analogy to Eq.~(\ref{isrm}). {Specifically, for an \textit{inter-system recurrence network (IRN)} constructed from two coupled complex systems, we} can formally {define} 
{
\begin{equation}\label{adjacency}
\mathbf{A}({\varepsilon})=\mathbf{IR}({\varepsilon})-\mathbb{I}_N,
\end{equation} }
\noindent
where $\mathbb{I}_N$ is the $N$-dimensional identity matrix {($N=N_X+N_Y$)}. In the resulting undirected and {unweighted} simple (no self-loops, multiple edges, or edge and vertex weights) network, the vertices represent the state vectors of {both} systems. Due to the block structure of {Eq.~(\ref{isrm})}, the IRN {consists of (i) two} (unipartite) RNs of the individual systems, the properties of which have been exhaustively studied elsewhere~{\cite{Marwan2009,Donner2010NJP,Donner2011IJBC,Donner2011EPJB,Donner2010PRE}}, and (ii) a bipartite {network that includes} only edges between vertices from different systems. In analogy to {CRPs}, we call the latter bipartite graphs  \textit{cross-recurrence networks}. 

{
In order to quantitatively study IRNs from a general coupled networks point of view~\cite{Donges2011EPJB} (see Section \ref{sec:measures}), the two subnetworks corresponding to the individual dynamical systems need to be clearly distinguished, i.e., there have to be only few cross-recurrences in comparison to intra-system recurrences ($RR^{XY}<RR^X,RR^Y$). Since this practical requirement is not always met by Eq.~(\ref{isrm}), it is necessary to modify the framework introduced above by allowing for different threshold distances $\varepsilon^X$, $\varepsilon^Y$ and $\varepsilon^{XY}$. Equations~(\ref{isrm}) and (\ref{adjacency}) thus become
\begin{equation}
\mathbf{IR}(\boldsymbol{\varepsilon})=\left( \begin{array}{cc} \mathbf{R}^X(\varepsilon^X) & \mathbf{CR}^{XY}(\varepsilon^{XY}) \\
\left[\mathbf{CR}^{XY}(\varepsilon^{XY})\right]^T & \mathbf{R}^Y(\varepsilon^Y) \end{array} \right) \label{isrm1}
\end{equation}
with
\begin{equation}
\boldsymbol{\varepsilon}=\left( \begin{array}{cc} \varepsilon^X & \varepsilon^{XY} \\ \varepsilon^{XY} & \varepsilon^Y \end{array} \right) \nonumber
\end{equation}
and
\begin{equation}\label{adjacency1}
\mathbf{A}(\boldsymbol{\varepsilon})=\mathbf{IR}(\boldsymbol{\varepsilon})-\mathbb{I}_N,
\end{equation} 
\noindent
respectively. In order to render the two individual RNs quantitatively comparable, it is recommended to choose $\varepsilon^X$ and $\varepsilon^Y$ independently such that $RR^X=RR^Y$~{\cite{Donges2011PRE,Donner2010PRE}}.

Since the IRN framework is based on purely geometric considerations,} the underlying time series do not need to have the same sampling. {Specifically, IRNs do} \textit{not} make use of any time information (which is distinctively different in comparison with many other existing methods of time series analysis) and exclusively {require} a reasonable attractor coverage in the (original or reconstructed) phase space by the sampled data.

\subsection{Characteristic measures for IRNs}\label{sec:measures}

The topological properties of general coupled networks can be quantified by a number of measures which represent a generalisation of the canonical measures of complex network theory~\cite{Donges2011EPJB}. In {the following}, we briefly review {those characteristics which are particularly important for characterising IRNs. Subsequently, we discuss how these measures} can indicate the direction of a coupling between certain dynamical systems. For a comprehensive study and detailed background of the underlying analytical theory, we refer to {\cite{Donges2011PRE,Donges2011EPJB}}.

\begin{figure}[ht]
 \centering
 \includegraphics[width=0.9\columnwidth]{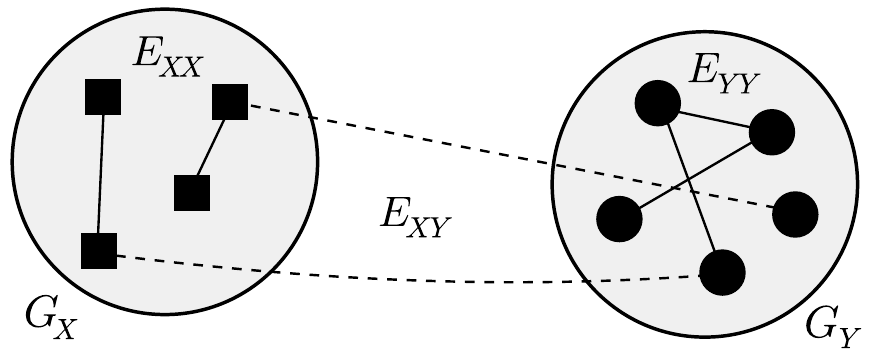}
 \caption{A possible representation of {two mutually coupled} networks. Solid lines mark internal edges (of edge sets {$E_{XX}, E_{YY}$}) while dashed lines represent cross-edges (elements of {$E_{XY}$}) connecting vertices {of both} subnetworks.}
 \label{fig:subnetworks}
\end{figure}

\begin{figure}[ht]
 \centering
 \includegraphics[width=0.95\columnwidth]{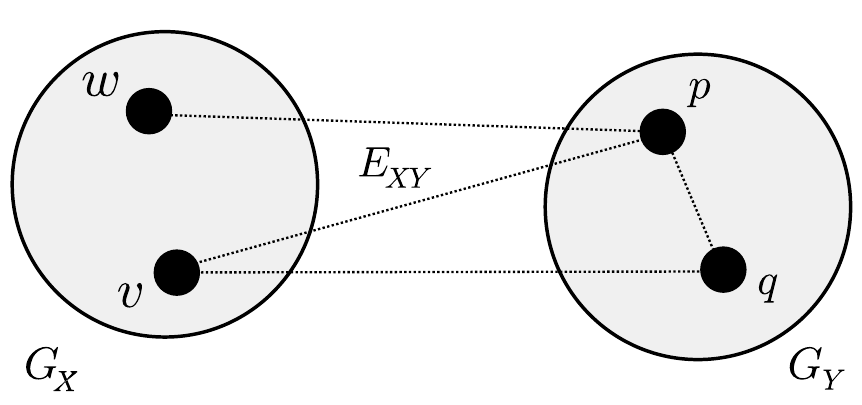}
 \caption{Two coupled subnetworks. The graph has global cross-clustering coefficients of {$\mathcal{C}^{XY}=0.5 \neq \mathcal{C}^{YX}=0$} and cross-transitivities {$\mathcal{T}^{XY}=1 \neq \mathcal{T}^{YX}=0$}.}
 \label{fig:interacting}
\end{figure}

Given a general undirected and unweighted simple graph $G=(V,E)$ described by the adjacency matrix $\textbf{A}$, consider a partition of the vertex set $V$ into {two} disjunct subsets {$V_X,V_Y \subseteq V$} such that {$V_X\cup V_Y = V$ and $V_X \cap V_Y = \emptyset$}. Similarly, the edge set $E$ is decomposed into disjunct {sets $E_{XX},E_{YY},E_{XY} \subseteq E$ with $E_{XX}\cup E_{YY}\cup E_{XY}= E$ and $E_{XX} \cap E_{YY}=\emptyset$, $E_{XX} \cap E_{XY}=\emptyset$, $E_{YY} \cap E_{XY}=\emptyset$} such that {$G_X=(V_X,E_{XX})$ and $G_Y=(V_Y,E_{YY})$ are} the induced {subgraphs} of the vertex {sets $V_X$ and $V_Y$} with respect to the full graph $G$ (Fig.~\ref{fig:subnetworks}). Then {$E_{XX}$} contains the (internal) edges within the subgraph or subnetwork {$G_X$ (likewise $E_{YY}$ those of $G_Y$)}, while {$E_{XY}$} comprises \mbox{(cross-)} edges {interconnecting the} subnetworks {$G_X$ and $G_Y$}. Furthermore, we can formally define bipartite subgraphs {$G_{XY}=(V_X\cup V_Y,E_{XY})$} containing all vertices of the vertex subsets {$V_X$ and $V_Y$}, and the (cross-) edges between these two sets. In the specific case of IRNs, the {$G_X$ and $G_Y$} correspond to the intra-system RNs constructed from the systems {$X$ and $Y$}, whereas {$G_{XY}$ contains} the cross-recurrence structure in terms of the sets of cross-edges $E_{XY}$. {In the following,} the letters $v,w,p,q$ shall {denote} single vertices (Figs. \ref{fig:subnetworks} and \ref{fig:interacting}).

{For a given vertex $v\in V_X$,} the \textit{cross-degree} 
{
\begin{equation}
k_v^{XY}=\sum_{q\in V_Y}A_{vq}
\end{equation}}
\noindent
gives the number of edges which connect $v$ {with} any vertex in subgraph {$G_Y$}. Taking the average over all {$v\in V_X$} and normalising yields the \textit{cross-edge density}
{
\begin{equation}
\rho^{XY}=\frac{1}{N_XN_Y}\sum_{p\in V_X,q\in V_Y} A_{pq}
\end{equation}}
\noindent
between the subgraphs {$G_X$ and $G_Y$}. For an IRN, {$\rho^{XY}$} equals the cross-recurrence rate {$RR^{XY}$} between the systems {$X$ and $Y$}.

The \textit{local cross-clustering coefficient} 
{
\begin{equation}
\mathcal{C}_v^{XY}=\frac{1}{k_v^{XY}(k_v^{XY}-1)}\sum_{p,q\in V_Y}A_{vp}A_{pq}A_{qv}
\end{equation}}
\noindent
estimates the probability that two randomly drawn neighbours of vertex {$v\in V_X$} from subgraph {$G_Y$} are also neighbours. For vertices $v^*$ with a cross-degree of zero or one, we define {$\mathcal{C}_{v^*}^{XY}=0$} in order to avoid divergencies. {By} averaging over all vertices {$v\in V_X$}, we {obtain} the \textit{global cross-clustering coefficient} {
\begin{equation}
\mathcal{C}^{XY}=\langle \mathcal{C}_v^{XY}\rangle_{v\in V_X}.
\end{equation}}
Closely related to {$\mathcal{C}^{XY}$} is the \textit{cross-transitivity}
{
\begin{equation}
\mathcal{T}^{XY}=\frac{\sum_{v\in V_X;p,q\in V_Y}A_{vp}A_{pq}A_{qv}}{\sum_{v\in V_X;p,q\in V_Y}A_{vp}A_{vq}},
\end{equation}}
\noindent
which, as an analog to the canonical network transitivity~\cite{Barrat2000,Newman2001,Newman2003}, counts the number of ``cross-triangles'' over the number of ``cross-triples''. It is important to note that both the cross-transitivity and the global cross-clustering coefficient are \textit{not} invariant under the permutation {$X\leftrightarrow Y$}, i.e., {$\mathcal{T}^{XY}\neq \mathcal{T}^{YX}$ and $\mathcal{C}^{XY}\neq \mathcal{C}^{YX}$} (Fig.~\ref{fig:interacting}), which is in fact the foundation of the method presented in the following.

\subsection{Network signatures of coupling}\label{sec:effects}

{For traditional RNs,} the canonical network measure of transitivity as well as the local and global versions of the clustering coefficient have already been recognised as valuable {characteristics} \cite{Marwan2009,Donner2010NJP,Zou2010,Donges2011PNAS}. Specifically, local clustering coefficient and global network transitivity can be interpreted as measures for the effective local and global dimension of chaotic attractors, respectively~\cite{Donner2011EPJB}. Following and expanding {these} ideas, we now take a look at the {``bivariate''} versions of these measures and how they behave under certain imposed constraints, in our case exemplified by a variable coupling between {two dynamical systems $X$ and $Y$}. 

\begin{figure}[bbb]
 \centering
 \includegraphics{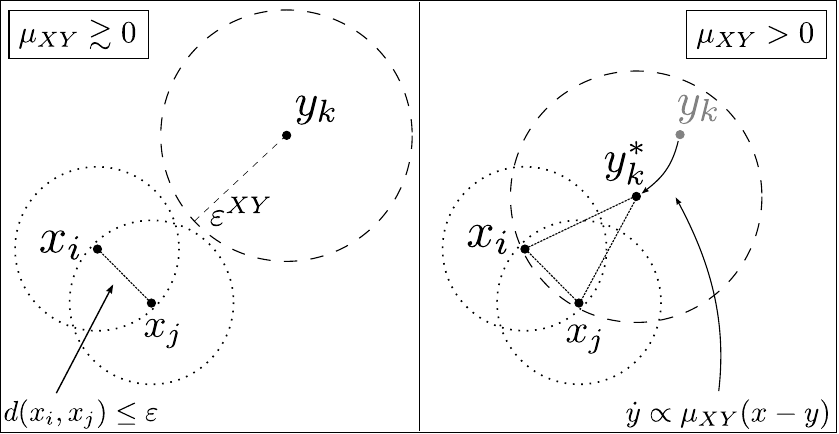}
 \caption{Schematic illustration of the influence of a diffusive coupling on the IRN structure. In the coupled case (right panel), the trajectory of $Y$ is dragged closer to $X$, while $X$ itself remains unchanged. We thus get an additional contribution to $\mathcal{C}^{YX}$ and $\mathcal{T}^{YX}$ in form of a ``cross-triangle''.}
 \label{fig:coupling}
\end{figure}

By taking into account both the univariate and the bivariate recurrence properties of the coupled system, {cross-transitivity and global cross-clustering coefficient are} unique in the way in which they act as a filter for information provided not by the individual systems' intrinsic processes, but by the way in which they are coupled. {Specifically, if} the two systems are sufficiently similar with respect to their invariant density, the geometric effects of coupling can be understood as follows (Table~\ref{tab:coupling}):

\begin{enumerate}[(i)]

\item In the \emph{uncoupled} case, we can consider $\mathcal{T}^{XY}$, $\mathcal{T}^{YX}$, $\mathcal{C}^{XY}$ and $\mathcal{C}^{YX}$ as randomly arising from the invariant densities of $X$ and $Y$ without any additional structural component. Therefore, we can expect $\mathcal{T}^{XY}\approx\mathcal{T}^{YX}$ and $\mathcal{C}^{XY}\approx\mathcal{C}^{YX}$.

\item Now consider a \emph{unidirectional} coupling of the type $\dot{y}\propto f(x-y)$, where $f$ is a monotonic function of either positive or negative sign (for example, in the important case of a diffusive coupling, $f(x-y)=\mu_{XY}(x-y)$, with $\mu_{XY}$ being the coupling strength). The strength and directionality of this coupling can be varied. Let $x_i$ and $x_j$ be two recurrent states in $X$. If the coupling direction is $X\to Y$ and the coupling is large enough, we are likely to also find a state $y_k^*$ in $Y$, which is (cross-) recurrent to both $x_i$ and $x_j$, due to the coupling's diffusive nature and thus the tendency to ``drag'' the trajectory of $Y$ towards $X$ (Fig.~\ref{fig:coupling}). The resulting ``cross-triangle'' adds to the value of both $\mathcal{T}^{YX}$ and $\mathcal{C}^{YX}$ according to their definition. On the other hand, ``cross-triangles'' constituted by two recurrent states in $Y$ and one cross-recurrent state in $X$ are merely coincidental due to the driver-response-like coupling. We would thus expect to see $\mathcal{T}^{YX} > \mathcal{T}^{XY}$ and $\mathcal{C}^{YX} > \mathcal{C}^{XY}$ in case of a unidirectional coupling $X\to Y$ and vice versa for the opposite coupling direction.

To put it differently: From the viewpoint of the driven system $Y$, the driving system $X$ ``contracts'' more than $Y$ from the perspective of $X$. Because $X$ is in fact unchanged (unidirectional coupling), this effect \textit{must} be due to changes of the spatial distribution of states in $Y$ induced by the coupling. This can be understood by looking at the generic form of the governing dynamical equations, $\dot{y}\propto f(x-y)$, i.e., the driven system $Y$ tends to minimise its distance to the driver $X$. 

\item Finally, for a \emph{symmetric bidirectional} coupling, the mutual effects on both systems are {comparable} and thus lead to IRN measures of the same magnitude. The same observation should hold if the subsystems become synchronised (at least in a generalised sense), e.g., in case of a sufficiently strong coupling. 

\end{enumerate}

\begin{table}
\begin{tabular}{c|c}
\hline
Coupling direction	 &	 Expected relation in network measures\\
\hline
no coupling & $\mathcal{T}^{XY} \approx \mathcal{T}^{YX}$, $\mathcal{C}^{XY} \approx \mathcal{C}^{YX}$\\
$X\rightarrow Y$ & $\mathcal{T}^{XY} < \mathcal{T}^{YX}$, $\mathcal{C}^{XY} < \mathcal{C}^{YX}$\\
$Y\rightarrow X$ & $\mathcal{T}^{XY} > \mathcal{T}^{YX}$, $\mathcal{C}^{XY} > \mathcal{C}^{YX}$\\
$X\leftrightarrow Y$ & $\mathcal{T}^{XY} \approx \mathcal{T}^{YX}$, $\mathcal{C}^{XY} \approx \mathcal{C}^{YX}$\\
\hline
\end{tabular}
\caption{Expected qualitative behaviour of IRN measures in different coupling situations for systems with comparable properties in the absence of (generalised) synchronisation.}
\label{tab:coupling}
\end{table}

{The above considerations are still mainly qualitative and still need to be supported by a detailed analytical theory of the geometric effects of coupling on the trajectories of dynamical systems. Specifically, the heuristic explanations presented for the (rather wide-spread) special case of diffusive coupling have to be further refined, and their extension to other types of coupling (e.g., impulse-coupling) has to be proven. Both aspects provide important directions for further research that are, however, beyond the scope of this work.}

\section{Example: Coupled chaotic oscillators}

{In the following we present a numerical analysis of the effect of coupling on IRN measures for a paradigmatic and well-studied model system. Specifically,} we demonstrate how our method performs in the detection of the coupling direction {between} two R\"ossler oscillators~\cite{Roessler1976} that are diffusively coupled via their second component:
\begin{align}\label{eq:roessler}
  \nonumber \dot{x^{(1)}}&=-(1+\nu)x^{(2)}-x^{(3)} \\
  \nonumber \dot{x^{(2)}}&=(1+\nu)x^{(1)}+a_X x^{(2)}+\mu_{YX}(y^{(2)}-x^{(2)})\\
  \nonumber \dot{x^{(3)}}&=b_X+x^{(3)}(x^{(1)}-c_X)\\
            \dot{y^{(1)}}&=-(1-\nu)y^{(2)}-y^{(3)} \\
  \nonumber \dot{y^{(2)}}&=(1-\nu)y^{(1)}+a_Y y^{(2)}+\mu_{XY}(x^{(2)}-y^{(2)}) \\
  \nonumber \dot{y^{(3)}}&=b_Y+y^{(3)}(y^{(1)}-c_Y),
\end{align}
where the parameters $a_{X,Y},\ b_{X,Y}$ and $c_{X,Y}$ define the shape (or \textit{regime}) of the respective attractors of the systems $X=(x^{(1)},x^{(2)},x^{(3)})$ and $Y=(y^{(1)},y^{(2)},y^{(3)})$, $\nu$ provides the systems with a frequency mismatch, and $\mu_{YX},\ \mu_{XY}$ denote the respective coupling strengths. Note the diffusive nature of the coupling term: The larger the coupling strength, the stronger the ``dragging'' of the driven system's trajectory towards the one of the driver. We vary the respective coupling strengths in such a way that we get unidirectional (one zero, one nonzero) or symmetric bidirectional ($\mu_{xy}=\mu_{yx}$) coupling situations.

\begin{figure*}
\centering
\subfigure[\ Coupling: $X\to Y$]{
\includegraphics[width=0.3\textwidth]{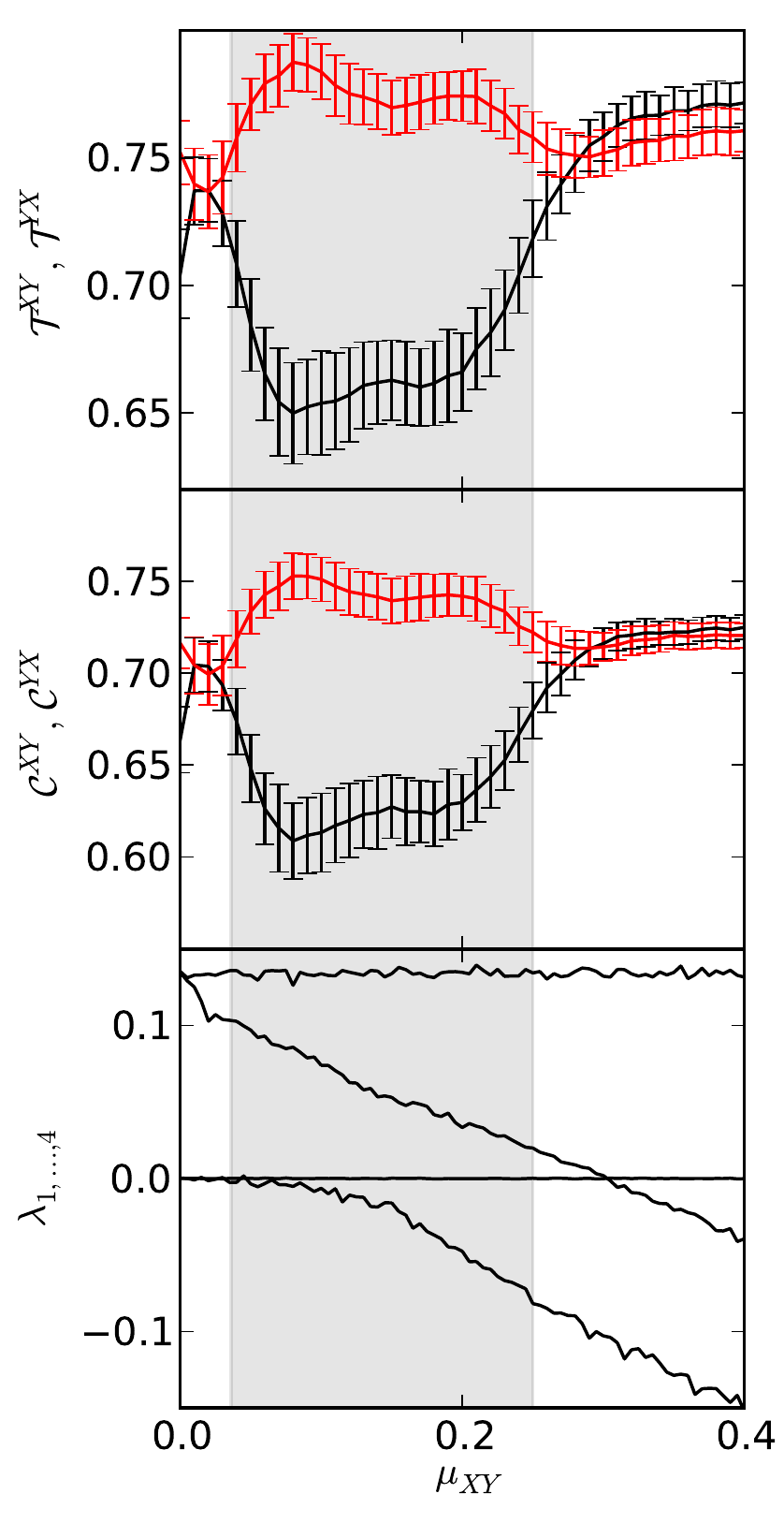}
}\label{fig:directionalFUN_XY} \hfill
\subfigure[\ Coupling: $Y\to X$]{
\includegraphics[width=0.3\textwidth]{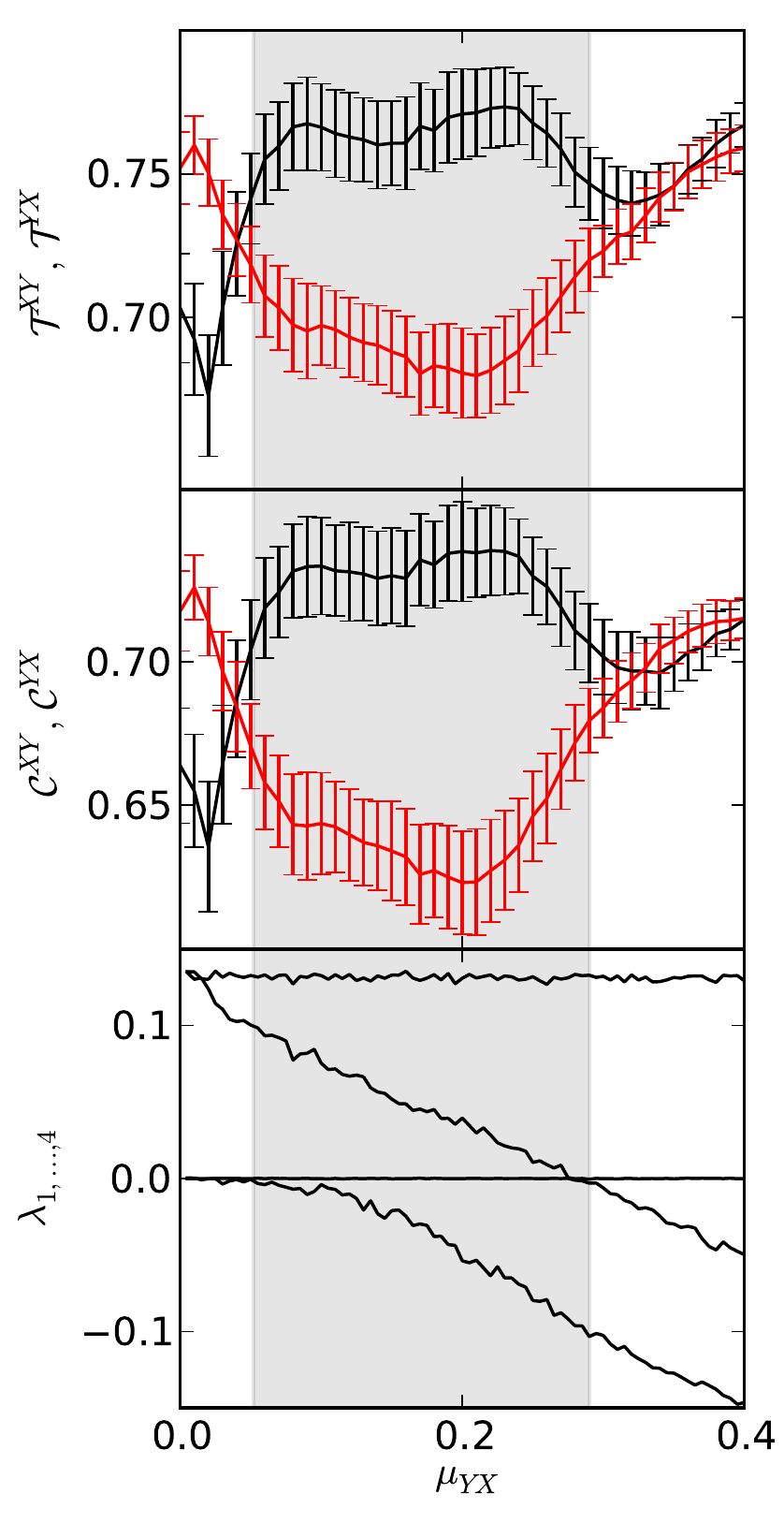}
}\label{fig:directionalFUN_YX} \hfill
\subfigure[\ Coupling: $X\leftrightarrow Y$]{
\includegraphics[width=0.3\textwidth]{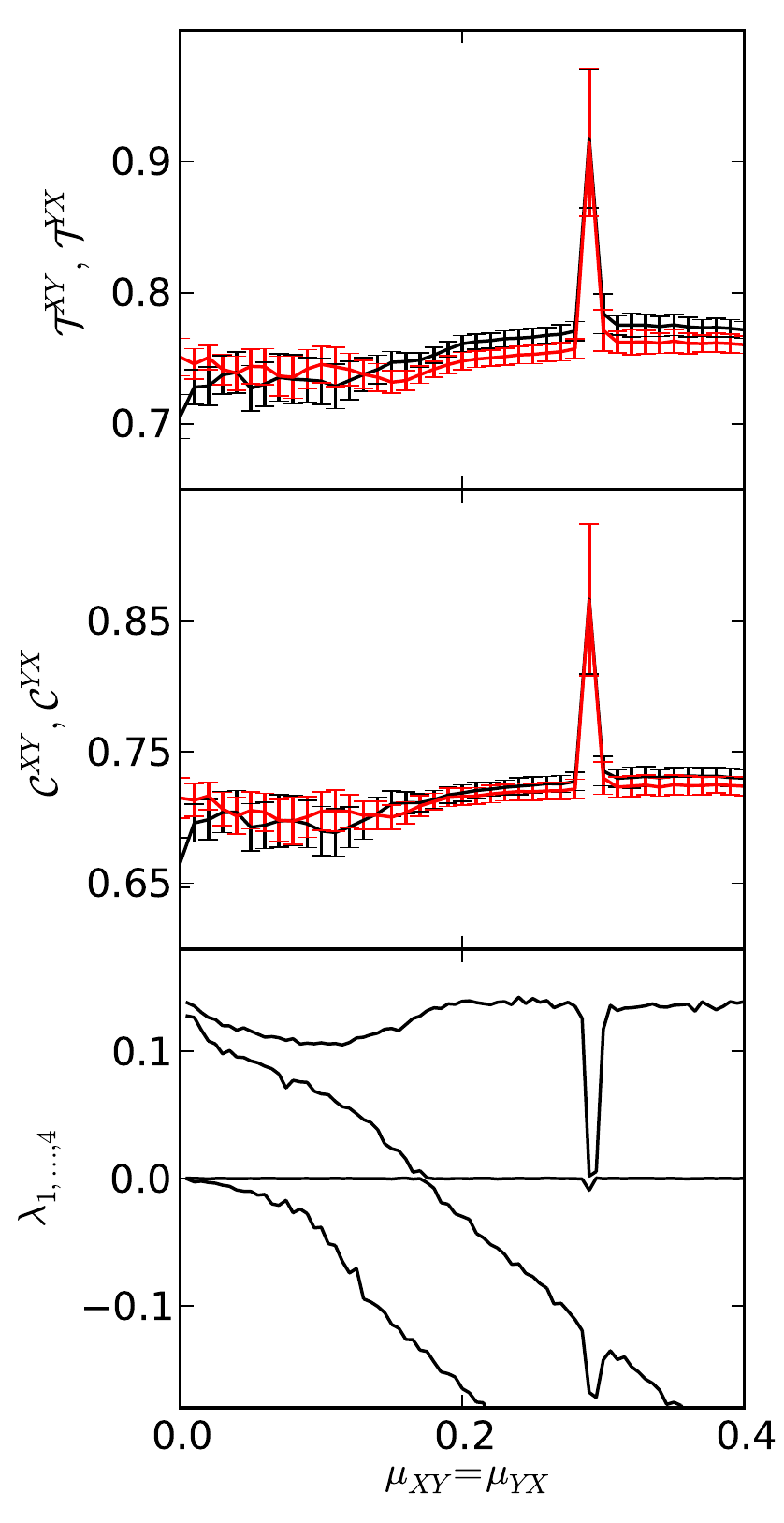}
}\label{fig:directionalFUN_BIDIR}
\caption{(Colour online) Coupling analysis for the R\"ossler system (Eq.~(\ref{eq:roessler})) in the funnel regime: Ensemble means and standard deviations (error bars) of the cross-network measures $\mathcal{C}^{XY},~\mathcal{T}^{XY}$ (black) and $\mathcal{C}^{YX},~\mathcal{T}^{YX}$ (red) and the four largest Lyapunov exponents. The Lyapunov spectrum has been estimated using the Wolf algorithm~\cite{wolf1985}. In the unidirectional cases (a) and (b), the coupling direction is correctly indicated for medium coupling strengths (shaded regions) until the onset of generalised synchronisation. In the symmetric bidirectional case (c), the interacting network measures are almost indistinguishable.}
\label{fig:directionalFUN}
\end{figure*}

\begin{figure*}
\centering
\subfigure[\ Coupling: $X\to Y$]{
\includegraphics[width=0.3\textwidth]{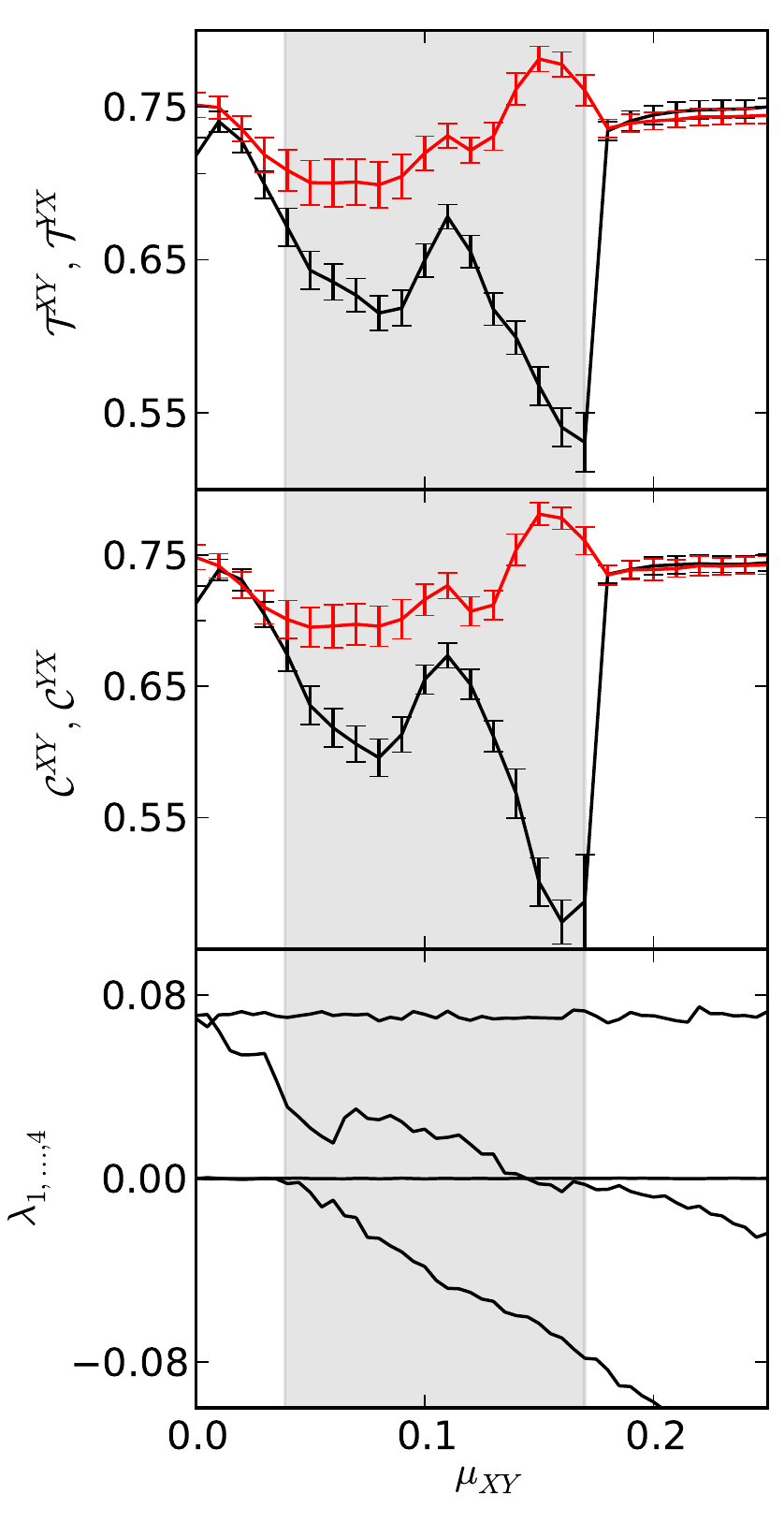}
}\label{fig:directionalPC_XY} \hfill
\subfigure[\ Coupling: $Y\to X$]{
\includegraphics[width=0.3\textwidth]{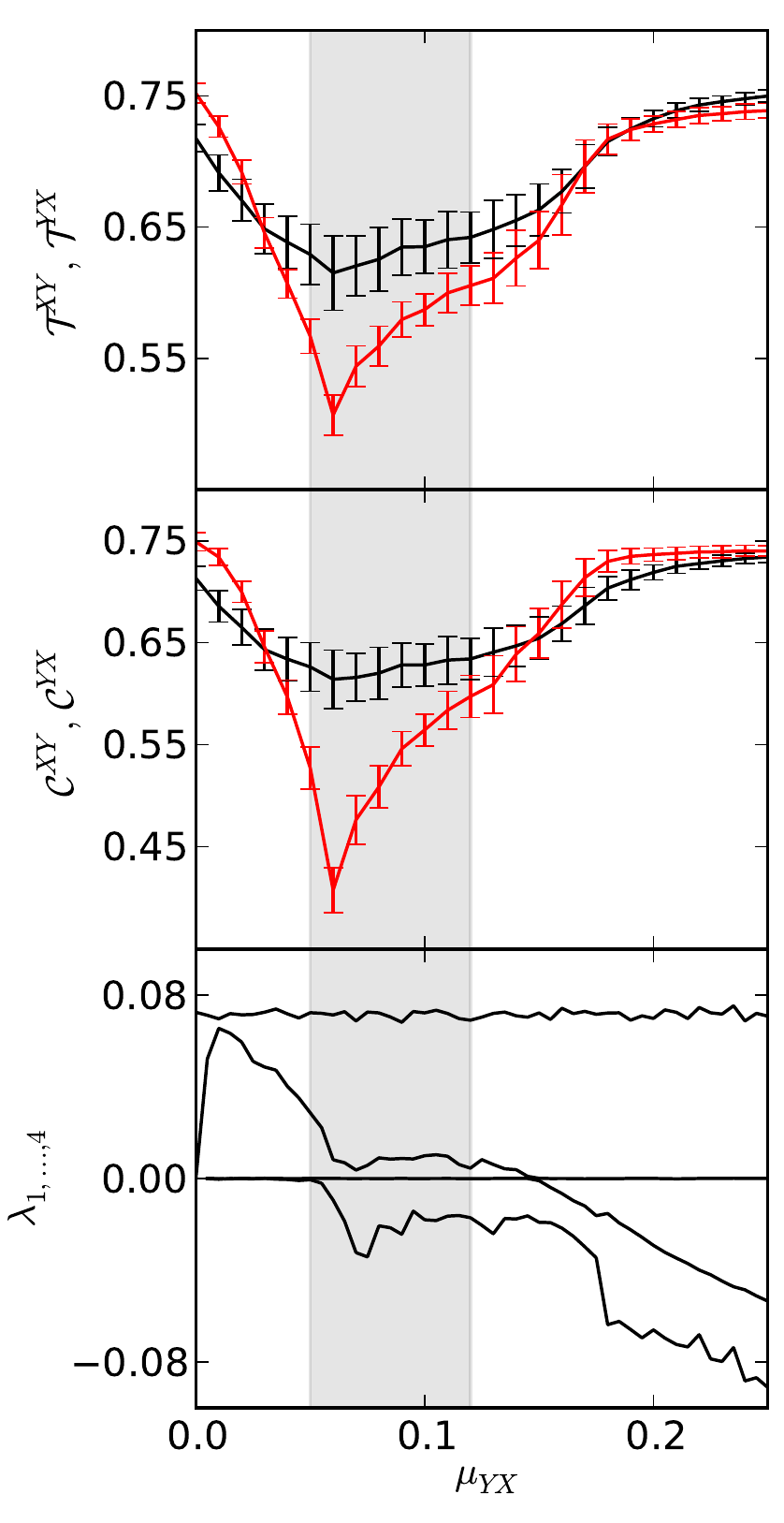}
}\label{fig:directionalPC_YX} \hfill
\subfigure[\ Coupling: $X\leftrightarrow Y$]{
\includegraphics[width=0.3\textwidth]{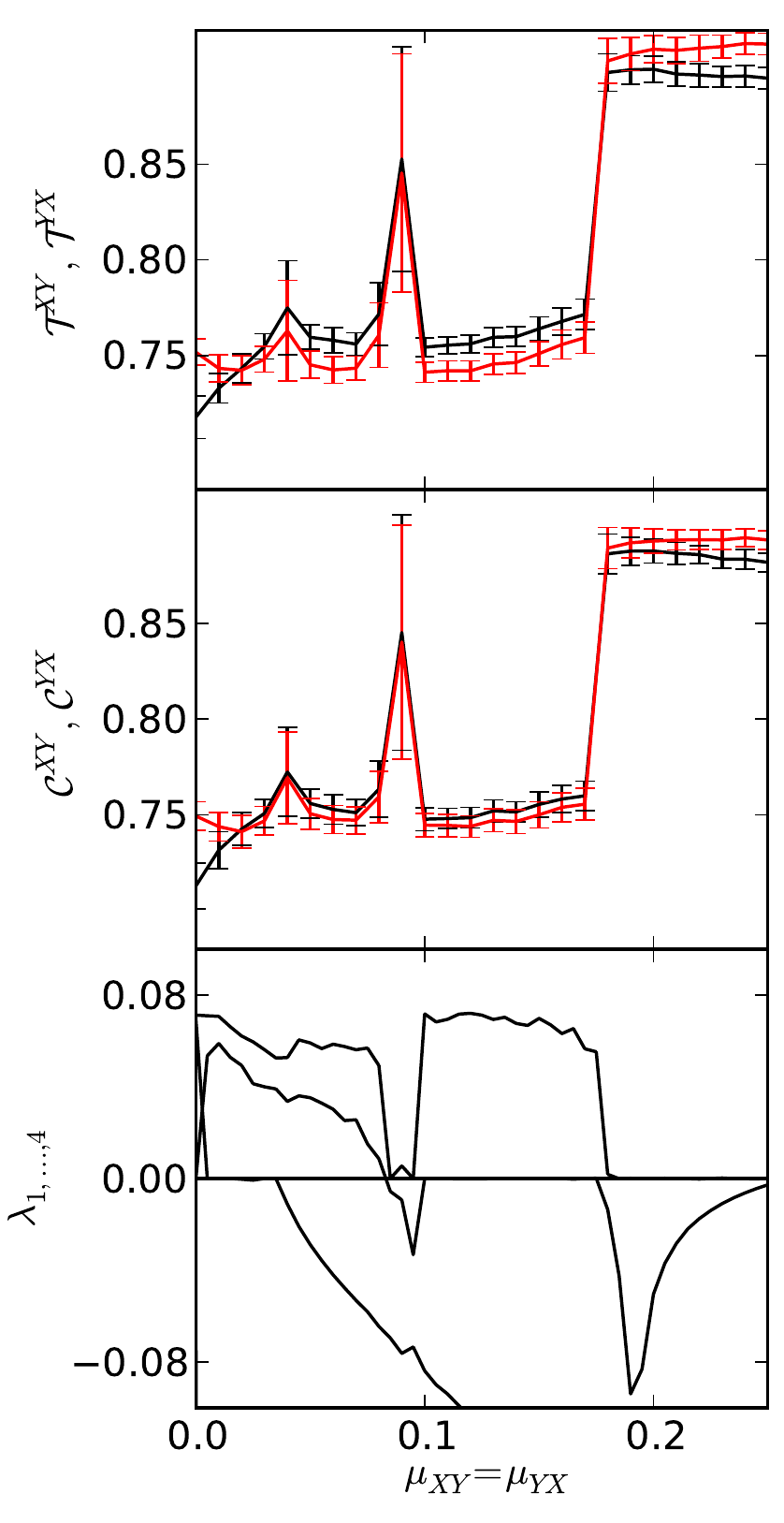}
}\label{fig:directionalPC_BIDIR}
\caption{(Colour online) As in Fig.~\ref{fig:directionalFUN} for the R\"ossler system in the phase-coherent regime.}
\label{fig:directionalPC}
\end{figure*}

For distinct values of the coupling strength, we create ensembles of $M=200$ realisations of the considered system by numerically \cite{hindmarsh1983odepack} integrating Eq.~(\ref{eq:roessler}) with a step size of $h=0.01$ for a total time $T=5,000$ using random initial conditions $x_0,y_0 \in [0,1]\times[0,1]\times[0,1]$. This ensemble approach was chosen to investigate the variability of the network measures of interest and ensure the reproducability of the results. Of the resulting $500,000$ points on each simulated trajectory, we discard the first $100,000$ as a possible transient phase, and then randomly choose $N=1,500$ points as a sample of state vectors from the attractor -- a technique which is widely known as bootstrapping \cite{Efron1979}. For both subsystems, we then infer $M=200$ univariate recurrence matrices $\mathbf{R}^X(\varepsilon^X)$ and $\mathbf{R}^Y(\varepsilon^Y)$ according to Eq.~(\ref{def:epsrec}), each with fixed thresholds $\varepsilon^X,\ \varepsilon^Y$ such that the recurrence rates $RR^X=RR^Y=0.03$, along with $M=200$ cross-recurrence matrices $\mathbf{CR}^{XY}(\varepsilon^{XY})$ with $\varepsilon^{XY}=\varepsilon^{YX}$ such that the cross-recurrence rate $RR^{XY}=0.02$. These choices may seem somewhat ad-hoc, but are consistent with results from previous works {\cite{marwan2007,Donner2010PRE,Schinkel2008,marwan2011}} suggesting recurrence rates $RR^X,RR^Y<0.05$ and considering that the coupled networks framework is only meaningful when $\rho^{XY}<\rho^X,\rho^Y$ (i.e., if the entire network has a pronounced modular structure such that it can be truly considered a network of networks).

\subsection{Inferring the coupling direction}\label{sec:coupldirection}

We follow the procedure described above for studying two different regimes of the R\"ossler system: the phase-coherent regime with
\begin{equation*}
a_{X,Y}=0.16,\ b_{X,Y}=0.1,\ c_{X,Y}=8.5,
\end{equation*}
and the non-phase coherent ``funnel regime'' with
\begin{equation*}
a_{X,Y}=0.2925,\ b_{X,Y}=0.1,\ c_{X,Y}=8.5.
\end{equation*}
In both parameter regimes, a frequency mismatch $\nu=0.02$ is considered to keep the otherwise identical subsystems from completely synchronising at very low coupling strengths, a case in which the (sub-)network structures would become fairly identical and an identification of the coupling direction by such means would be impossible. In case of unidirectional coupling (Figs.~\ref{fig:directionalFUN} and \ref{fig:directionalPC}), we find that for a large interval of intermediate coupling strengths, the coupling direction is identified correctly according to our hypothesis: For a coupling direction $X\to Y$ ($\mu_{XY}>0,\,\mu_{YX}=0$), we notice that the ensemble means and standard deviations of $\mathcal{C}^{XY}$ and $\mathcal{T}^{XY}$ are clearly below those of $\mathcal{C}^{YX}$ and $\mathcal{T}^{YX}$, respectively, and vice versa for $Y\to X$. {In general, the results obtained for both measures are qualitatively similar. However, although they are conceptually related, both quantify distinctively different properties of the networks under study, i.e., the observed similarity is no trivial result, but reflects the robustness of the geometric modifications to the driven system due to the imposed coupling.}

For small coupling strengths, a discrimination is not possible, since the geometric signatures of coupling are too weak to be resolved by our network analysis. For stronger couplings, the cross-network measures become equal again. We attribute this to the occurence of generalised synchronisation, which is indicated by the zero-crossing of the second-largest Lyapunov exponent (Figs.~\ref{fig:directionalFUN} and \ref{fig:directionalPC}). This observation is consistent with the case of bidirectional coupling, where the cross-network measures remain equal (Figs.~\ref{fig:directionalFUN}(c) and \ref{fig:directionalPC}(c)).

\subsection{The onset of synchronisation}

\begin{figure}
\centering
\includegraphics[width=\columnwidth]{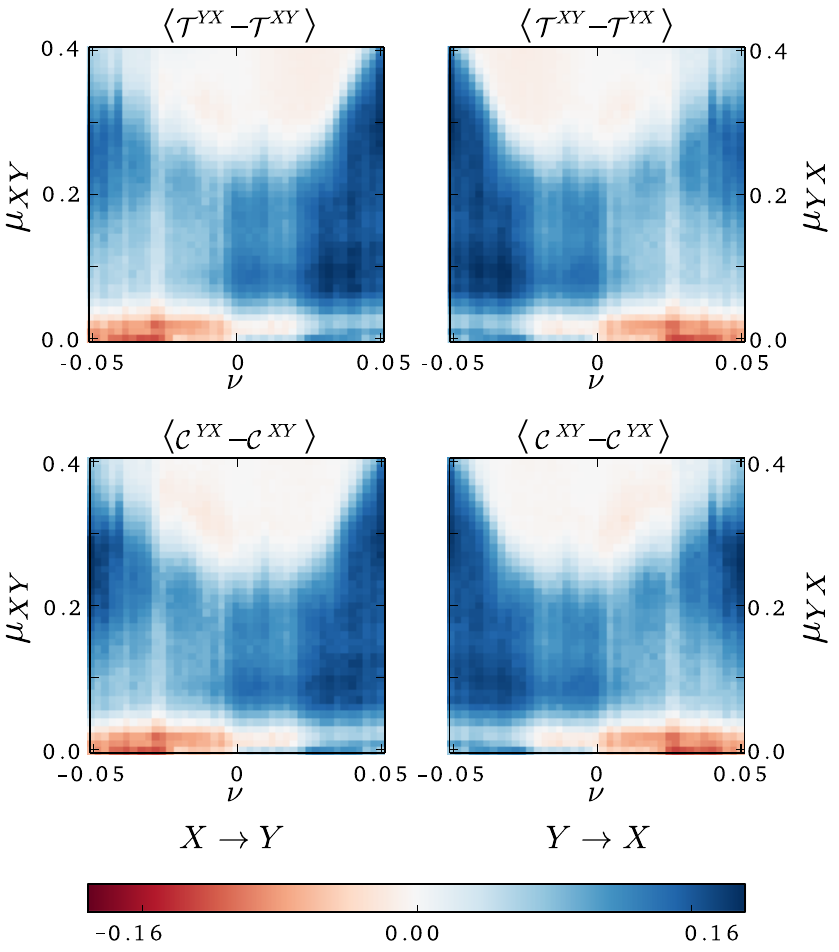}
\caption{(Colour online) Differences between the cross-transitivities (top) and global cross-clustering coefficients (bottom) for two coupled R\"ossler systems in the funnel regime for $X\to Y$ (left) and $Y\to X$ (right) coupling. The well-expressed Arnold tongue (white) indicates a parameter region with generalised synchronisation.}
\label{fig:arnold}
\end{figure}

Next, we further investigate the conditions related to the emergence of generalised synchronisation and, thus, the breakdown of our method. For this purpose, we expand our analysis into a two-dimensional parameter space, varying not only the coupling strength, but also the system's frequency mismatch $\nu$. The considered discrete $(\nu,\mu)$-space consists of $41\times 41=1,681$ points (Fig.~\ref{fig:arnold}) . For each of these points, we generate $M=100$ IRNs and calculate the mean differences of the network measures $\left<\mathcal{T}^{XY}-\mathcal{T}^{YX}\right>$ and $\left<\mathcal{C}^{XY}-\mathcal{C}^{YX}\right>$, and vice versa. 

At low frequency mismatches, the systems synchronise at lower coupling strengths, leading to a balancing of the cross-network measures (i.e., differences between the measures vanish) and thus a failure of the detection of the coupling direction already at relatively low values of the coupling strength. {As a result,} Fig.~\ref{fig:arnold} features white regions, indicating zero difference in the measures corresponding to the well-known Arnold tongues and revealing the onset of generalised synchronisation  \cite{Pikovsky_Kurths_synchr}. This was anticipated and retrospectively justifies our corresponding choice of a nonzero detuning as used above.

{Moreover, in Fig.~\ref{fig:arnold}} positive values indicate {parameters} in which the coupling direction was correctly identified (blue regions). In turn, negative values
correspond to a wrongly detected coupling direction (red regions in Fig.~\ref{fig:arnold}), which appears only for very small coupling strength if the driven system has a higher frequency than the driver. This observation suggests that in the latter case, the considered detuning has a geometric effect on the two attractors that overcompensates the actual signatures of (weak) coupling.

\subsection{{Limits} of the proposed framework}

\begin{figure}
\centering
\includegraphics[width=0.75\columnwidth]{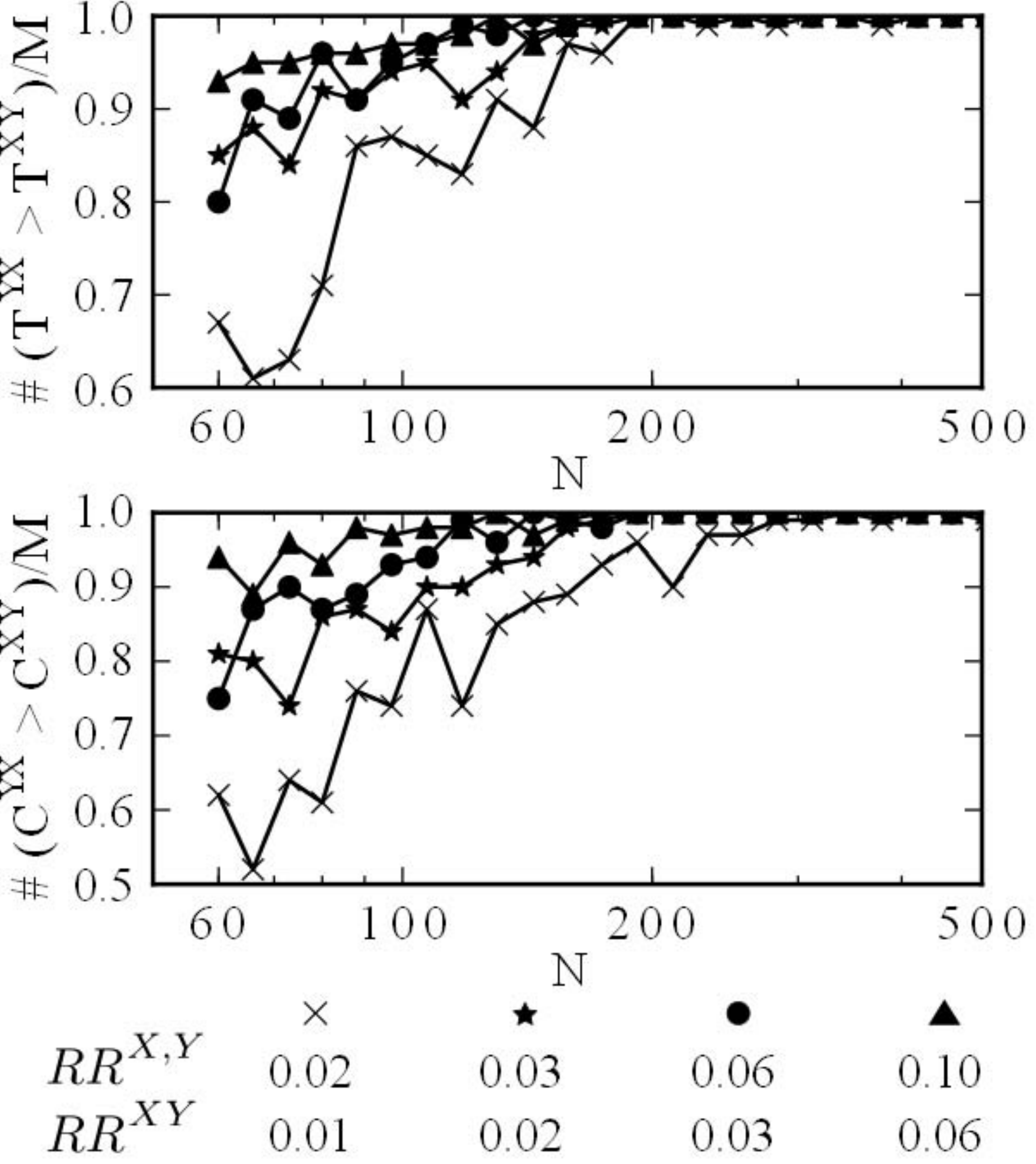}
\caption{Ratio of correct identifications of the coupling direction $X\to Y$ by means of cross-transitivity and global cross-clustering for the funnel regime of the R\"ossler system depending on the network size. Higher recurrence rates yield better distinctions. In all cases $N\geq 300$ is a sufficient lower boundary.}
\label{fig:nodes}
\end{figure}

To explore further properties and possible advantages of our method, we now strive to find a lower boundary for the number of vertices in the IRN necessary for a correct identification of the coupling direction. As an example, we again consider a unidirectional $X\to Y$ coupling between two R\"ossler oscillators in the funnel regime with $\mu_{XY}=0.1$ and a frequency mismatch of $\nu=0.02$. In {Section~\ref{sec:coupldirection}} the chosen number of $N=1,500$ vertices ensured a good discrimination. In order to find a lower bound for $N$ we vary the number of samples bootstrapped from the integrated trajectory and repeat our procedure again, each time calculating $M=100$ realisations of the trajectory with random initial conditions for generating our IRNs. In order to see whether or not a distinction is possible, we consider the number of correct and false identifications (Fig.~\ref{fig:nodes}). Moreover, the same procedure was repeated for different values of $RR^{X},\,RR^{Y}$, and $RR^{XY}$ to also test the robustness of the method with respect to a variation of the recurrence threshold (Fig.~\ref{fig:nodes}).

{In all cases examined here,} for a perfect discrimination (100 correct, 0 false cases), no more than roughly $N=300$ vertices {were} necessary. This is a remarkable advantage for real-world applications, where the amount of available data is naturally much more limited than in the case of model systems. The distinction is also robust with respect to variations of the recurrence threshold, with higher recurrence rates yielding better results {for shorter time series}.

\section{Real-world example: Palaeoclimate variability}

In order to demonstrate the potential of {IRNs} for the analysis of real-world time series, we make use of two proxy records of climate variability in the Asian monsoon system during the last about 10,000 years (Holocene). The Asian monsoon system is not only one of the most important atmospheric circulation systems, it also has a strong socio-economic impact because it affects a major part of the global population. The instability of the Asian monsoon has been correlated with agricultural and cultural prosperity and political unrest in the past \cite{Zhang2008Science}. Hence, understanding the mechanisms that lead to an interrupted, weakened, or enhanced Asian monsoon is of paramount importance in the on-going discussion on global climate change and the monsoon as a tipping element \cite{Lenton2008}. 

\begin{figure*}[t!]
\centering
\includegraphics[width=.8\textwidth]{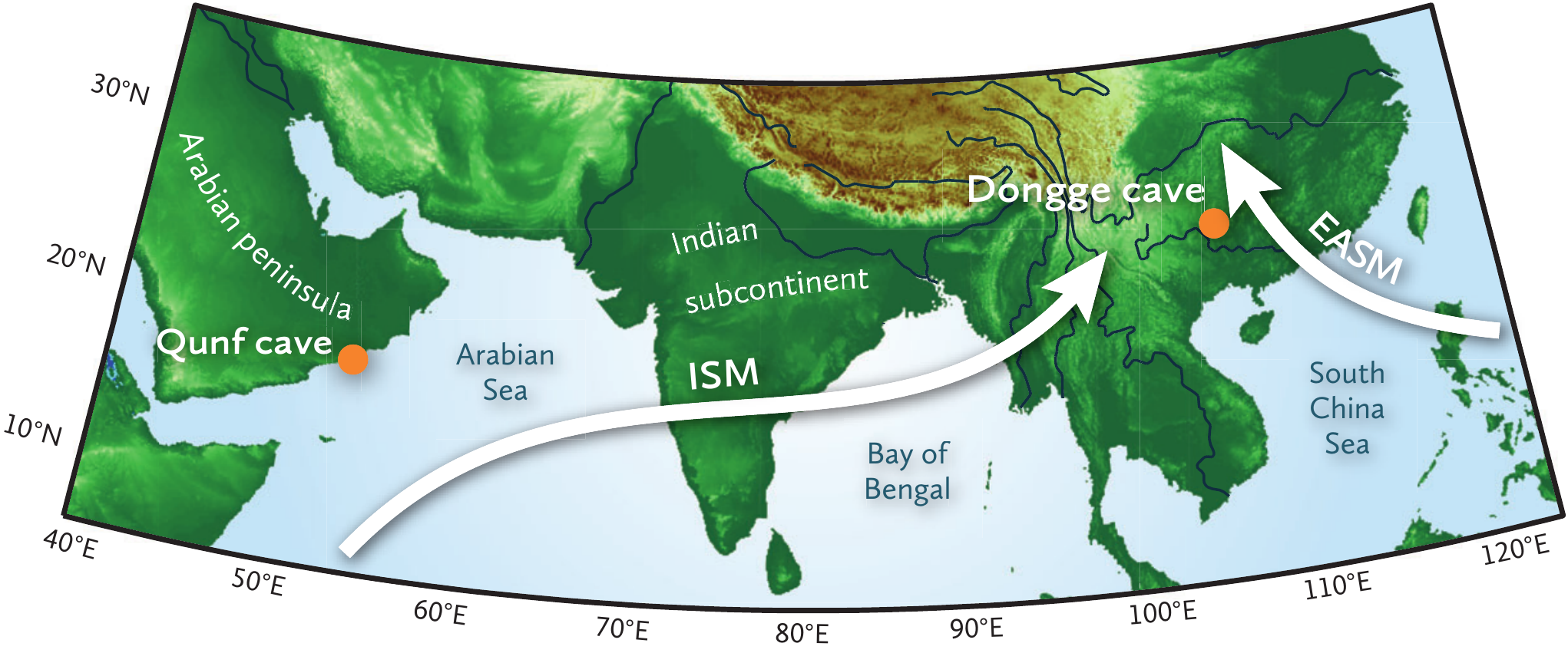}
\caption{(Colour online) Main wind directions of the Indian (ISM) and East-Asian summer monsoon (EASM) and locations of the Qunf and Dongge caves \cite{Liu2008,Clemens2010}.}
\label{fig:map}
\end{figure*}

The Asian summer monsoon can be mainly divided into the Indian and the East Asian monsoon subsystems (ISM and EASM), which transport moisture from {distinct} sources to the continent (Fig.~\ref{fig:map}). The study of interrelationships between the different monsoon subsystems contributes to a better understanding of the entire monsoon system. Variations in oxygen isotope ratios {($\delta^{18}$O)} measured in speleothems provide high-resolution proxies for monsoonal activity in the past \cite{Wang2005,Fleitmann2003}. Based on such proxy records, significant spatio-temporal changes in the {EASM} due to different global climate regimes have been recently found \cite{rehfeld2011}. However, the ISM is less well understood with respect to its long-term evolution, which is partially due to the lower number of available high-resolution palaeoclimate data from the Indian {subcontinent}.

\begin{figure}
\centering
\includegraphics[width=\columnwidth]{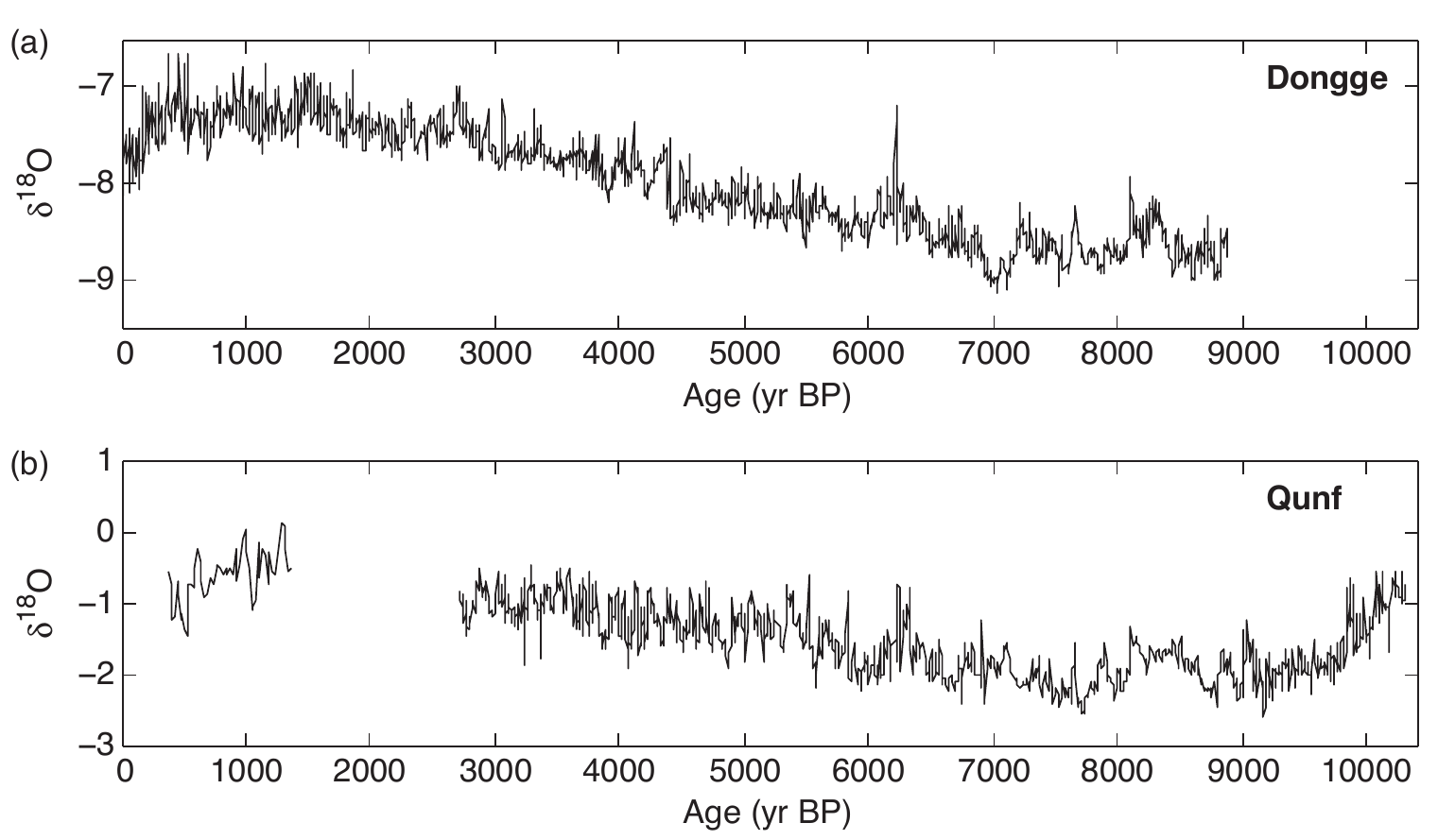}
\caption{Palaeoclimate proxy records of $\delta^{18}$O (see text for a detailed description) from speleothems sampled from the Dongge (a) and Qunf (b) caves.}
\label{fig:data}
\end{figure}

{Here}, we aim to study the prevalent direction of coupling between ISM and EASM since the end of the last glacial period. We use the variability of $\delta^{18}$O measured in stalagmites (a proxy for the strength of the summer monsoon) obtained from the Dongge cave in Southeast China \cite{Wang2005} and the Qunf cave in Oman \cite{Fleitmann2003} (Fig.~\ref{fig:map}). The Dongge record $D$ represents Holocene monsoon variations in the EASM region, while the Qunf record $Q$ contains information on monsoonal climate variability in the ISM region~(Fig.~\ref{fig:data}, Tab.~\ref{tab:data}). {We emphasise that there is no formal evidence that both data sets actually quantify exactly the same climate variable, since the same proxy could have been partially influenced by different variables in different ways according to the specific local conditions (e.g., evaporation, dripwater rate, CO$_2$ degassing, host rock, etc.). However, in agreement with the paleoclimate literature~\cite{Wang2005,Fleitmann2003}, we are confident that using the same type of proxy data from the same type of paleoclimate archive allows considering both records as representing approximately the same integral observable (specifically, the average annual amount of rainfall, which by itself stands as a proxy for the strength of the summer monsoon since most of the annual precipitation at both studied sites is in fact associated with this season). Moreover, since the global climate system is continuous in space, we have both time series representing the dynamics of different components of the same dynamical (physical) system. Consequently, we conjecture that the processes recorded in both time series are sufficiently interrelated and similar to allow for an application of the proposed IRN methodology.}

\begin{table*}
\caption{Time interval $T$ (BP $=$ years before 1950), mean sampling time $\Delta t$, number of observations $N$, and corresponding reference of the studied climate proxies.}\label{tab:data}
\begin{tabular}{lcccl}
\hline
Record	&$T$ (yr BP)	&$\Delta t$ (yr)	&$N$	&References\\
\hline
Dongge $D$	&-50\dots 8880	&4.2			&2,124	&Wang et al. (2005) \cite{Wang2005}\\
Qunf	$Q$	&378\dots 10,300	&7.1			&1,405	&Fleitmann et al. (2003) \cite{Fleitmann2003}\\
\hline
\end{tabular}
\end{table*}

For the two datasets $D$ and $Q$ we first apply detrending by subtracting a 100-yr moving-average from all data (for a detailed description of the corresponding procedure, see \cite{Donges2011NPG}). We emphasise that this is crucial for the interpretation of all further results of our analysis, since we restrict ourselves to {variability} on inter-annual to multi-decadal time scales (i.e., scales below 100 years). As a second step, we use time-delay embedding of the data with an embedding dimension of $d=3$ in both cases but different time delays of $\tau_{D}=3$ for the Dongge record and $\tau_{Q}=2$ for the Qunf record. These parameters have been chosen according to the results of the false nearest neighbors and auto-mutual information methods \cite{kantz1997}. The {distinct} time delays account for {the time series'} different average sampling rates, leading to $\tau\approx 13$ years on average for both records to make the considered state vectors actually comparable (cf.~Tab.~\ref{tab:data}). Note that for the embedding step, we neglect possible variations of the sampling rate with time, as has been done before in successful applications of univariate recurrence network analysis to palaeoclimate data \cite{Marwan2009,Donges2011NPG,Donges2011PNAS}. {This simplification possibly leads to modified positions of individual state vectors in the reconstructed phase space. However, since the exact timing of all data points is not known and the sampling rate does not change too much with time, we argue that within the uncertainty of both the recorded observable and the timing of measurements, the approximated low-dimensional attractor has a shape that does not differ markedly from that one would reconstruct when having ``perfect'' (equidistant and completely certain) observations.}

{In the considered example, the auto-correlations of both records decay sufficiently fast (i.e., fall below $1/e$ within one sampling time step on average). This allows us excluding severe effects due to ``false'' recurrences corresponding to subsequent observations, which can occur if the sampling is very dense in comparison with the typical time-scale on which the auto-correlations decay. In such cases, however, removing the corresponding sojourn points from the recurrence matrix by introducing a suitable Theiler window would provide a simple and feasible solution~\cite{Donner2010NJP}. Following these considerations, the requirement of a ``sufficient attractor coverage'' could be turned into more explicit conditions (such as time series covering at least about 10-20 oscillations of a chaotic system, or a comparable requirement on the total number of data for general real-world time series with a typical preferred time scale).}

{Having thus obtained time series in a reconstructed phase space that is quantitatively comparable for both records}, from the embedded state vectors we bootstrap 80\% of the data and calculate the resulting IRNs and their cross-transitivities and global cross-clustering coefficients, with $RR^D=RR^Q=0.03$ and $RR^{DQ}=RR^{QD}=0.02$ as in Section 3. This procedure is carried out $M=10,000$ times to evaluate the robustness of our results. {The chosen size of the bootstrap samples presents a trade-off between the requirements of a sufficiently large number of different samples and sufficiently large individual samples. We emphasise that because the IRN approach does not make any assumptions on the timing of observations, we can directly apply this method, including the considered bootstrapping approach, to the embedded time series disregarding their irregular sampling and imperfect chronological control. In fact, this is an important advantage of our network-based approach in comparison with RQA and similar techniques, which rely on line structures in a recurrence plot and, hence, uniform sampling.}

\begin{figure}
\centering
\includegraphics[width=\columnwidth]{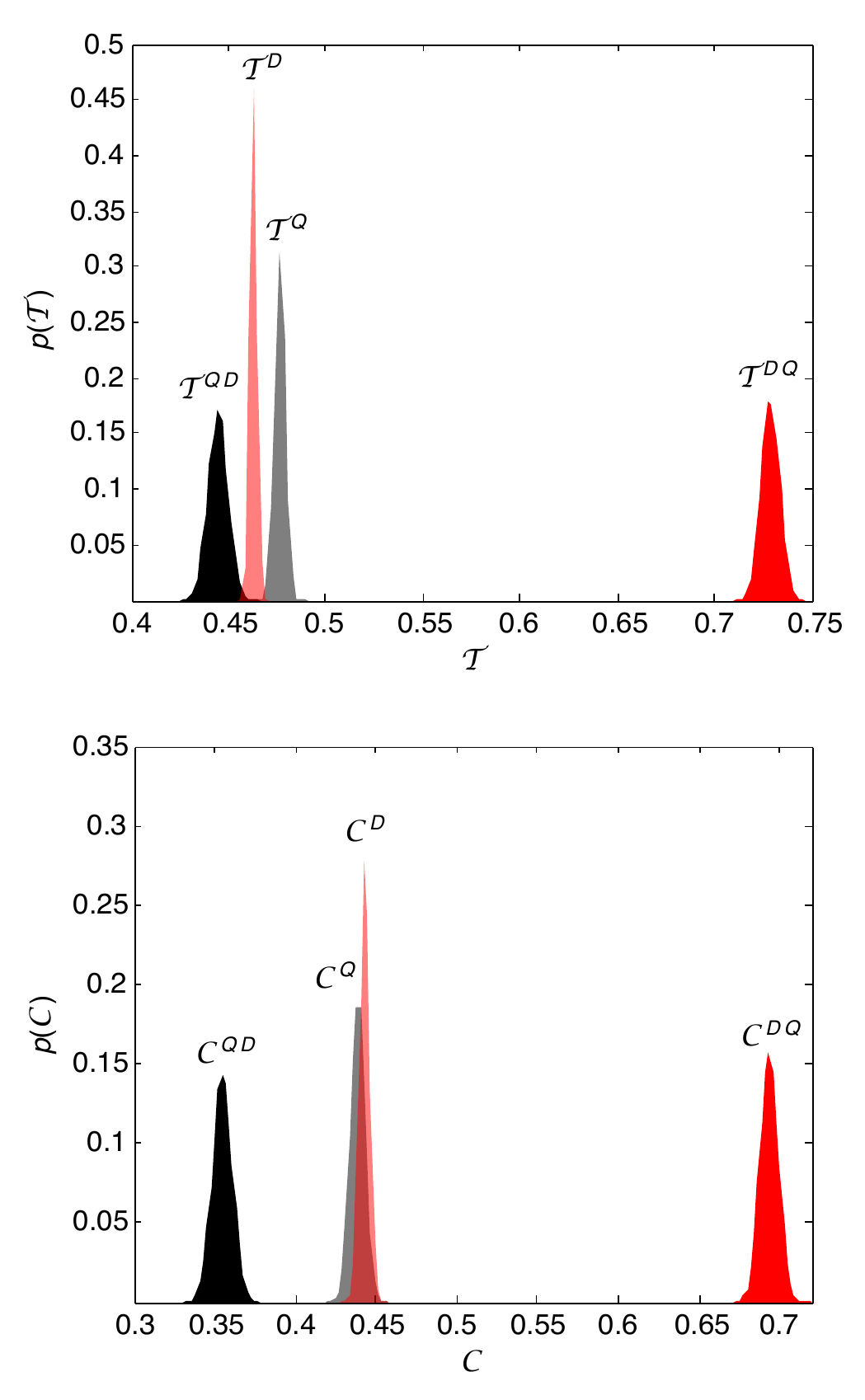}
\caption{(Colour online) Normalised distributions of the IRN measures cross-transitivity (top) and global cross-clustering coefficient (bottom) inferred from the Dongge and Qunf data by bootstrapping 80\% of the data in $M=10,000$ {independent} realisations. The distributions of the inter-system measures are clearly {separated} with $(\mathcal{C}^{DQ},\mathcal{T}^{DQ}>\mathcal{C}^{QD},\mathcal{T}^{QD})$, indicating a possible coupling direction $Q\to D$.}
\label{fig:data1}
\end{figure}

{As a result, we find} that the bootstrapped distributions of the {traditional RN} transitivities and global clustering coefficients of the two respective records display a considerable overlap (Fig.~\ref{fig:data1}). {This observation} indicates that the two climate subsystems {(}the dynamics of which {have} been recorded by the considered palaeoclimate archives{)} are sufficiently similar with respect to their long-term dynamics so that both can actually be viewed as sharing the same phase space. {The latter} finding motivates the further consideration of our IRN approach for detecting a possible directional coupling between both systems. In fact, the resulting distributions of cross-transitivities and global cross-clustering coefficients are clearly separated by $\mathcal{T}^{DQ}>\mathcal{T}^{QD}$ and $\mathcal{C}^{DQ}>\mathcal{C}^{QD}$. {This general observation persists when interpolating both records to a common and uniform sampling before constructing the IRN (not shown), underlining the feasibility and robustness of the considered approach. This suggests} that the coupling direction was $Q\to D$ during the Holocene. {Specifically,} under the {assumptions that (i)} $Q$ represents the ISM activity and $D$ the EASM dynamics, {(ii) auto-correlations do not play a significant role on the considered time-scales, and (iii) a possible coupling between both climate subsystems is of diffusive nature, we can thus argue} that the monsoon impact at the location of the Chinese cave was {probably} not only controlled by the EASM but also by the ISM, whereas the EASM did not have {a comparably strong} impact on the location of the Oman cave.

{A possible} directed influence of one monsoon branch to locations usually influenced by the other monsoon branch is of {considerable} importance for understanding the dynamical {mechanisms} of the monsoonal system (e.g., as proposed by \cite{pausata_ngeo2011}). The potential directionality of ISM towards the EASM region {indicated by our results} suggests that the ISM branch is probably stronger, at least on average during the considered period of time. This {could also imply} that the Indian ocean (in particular the Indian ocean sea surface temperature, the principal pattern of which is the Indian Ocean Dipole, IOD \cite{abram_ngeo2008}) plays not only an important regional role for the Indian subcontinent, but also for the entire Asian monsoon system. The EASM is strongly influenced by the El Ni\~no/Southern Oscillation (ENSO), whereas the ENSO impact on the monsoon rainfall in India is {known to be} more subtle and {changing} over time~\cite{Kumar1999,Maraun2005,Wu2012}. The detected directionality adds a new aspect to this relationship and might help in further improving our understanding of the Asian monsoon system as a whole. We particularly emphasise that IRNs can provide important information for constructing and interpreting networks of palaeoclimate records~\cite{rehfeld2011,Rehfeld2012} to infer dominating spatial patterns of climate dynamics in the Earth's history.

\section{Conclusions}

We have proposed a new {complex} network-based approach for inferring coupling directions from bivariate time series. The recently introduced concepts of recurrence networks \cite{Donner2011IJBC} and coupled network analysis \cite{Donges2011EPJB} have been combined to the new analytical tool of inter-system recurrence networks {(IRNs)}. Cross-transitivity and global cross-clustering coefficient {of IRNs} can be used to quantitatively derive the coupling direction {based on purely geometric considerations. Although developing a corresponding generally applicable theoretical understanding, as well as a systematic comparison with other existing techniques for detecting the coupling direction between two measured signals, will be subject of future research, our results suggest that the proposed framework can be applied to many situations where the exact functional form of coupling is not necessarily known.}

We have demonstrated the potential of the proposed framework using a prototypical model system and discussed its robustness and sensitivity with respect to data length and recurrence threshold. We find that the {IRN} method performs well even for short time series ($N\approx 300$). {Another important advantage of our approach is that it relies on the spatial structure of the coupled attractors in phase space and does not require any time-information. Hence, problems of strong auto-correlations often present in real-world systems are avoided by construction. In turn, our method is not able to detect possible coupling delays unless being applied separately for different, mutually shifted time slices.}

{The proposed approach can be directly generalised to studying mutual couplings among a set of $K>2$ subsystems, which leads to IRNs the adjacency matrices of which have a block structure composed of $K$ traditional (unipartite) recurrence networks and $K(K-1)/2$ bipartite cross-recurrence networks. Note, however, that the study of couplings for $K>2$ is much more challenging than for $K=2$, since in this case both direct and indirect interactions may be present and need to be distinguished. It will be subject of future research to investigate if and how the proposed geometric framework based on multivariate extensions of recurrence networks is able to cope with this problem.}

{As a real-world example}, we have studied the {possible} directionality of coupling between the Indian Summer Monsoon (ISM) and the East Asian Summer Monsoon (EASM) during the {last about 10,000 years}. Our method allows its direct application to the data without interpolation to a common time axis. From our analysis we {find indications that} the ISM is probably influencing the monsoon strength in Eastern China (located in the EASM influence region) on {multi-decadal} time scales between about 10 years (data resolution) and 100 years (scale of detrending), which supports previous findings based on modelling \cite{pausata_ngeo2011}. However, further research is necessary to unambiguously verify this result by means of complementary methods of bivariate time series analysis. 

{As a final remark, it is worth mentioning that there are other, conceptually different multivariate generalisations of recurrence plots, such as joint recurrence plots~\cite{Romano2004} and multivariate recurrence and cross-recurrence plots~\cite{Nichols2006}. In principle, all these concepts could be re-interpreted from a complex network point of view as well. However, due to their different construction mechanisms, they provide different types of information on the investigated systems in comparison with the approach presented in this work. A detailed corresponding study will be subject of future work.}\\

\noindent
\emph{Acknowledgments.}
This paper was partly developed within the scope of the IRTG 1740/ TRP 2011/50151-0, funded by the DFG/ FAPESP. Furthermore, this work has been financially supported by the Leibniz Society (project ECONS), the German National Academic Foundation, the Potsdam Research Cluster for Georisk Analysis, Environmental Change and Sustainability (PROGRESS, support code 03IS2191B), and the DFG research group ``Himalaya: Modern and Past Climates (HIMPAC)''. For calculations of complex network measures, the software package \texttt{pyunicorn} has been used on the IBM iDataPlex Cluster of the Potsdam Institute for Climate Impact Research. We thank Kira Rehfeld and Sebastian Breitenbach for helpful comments and discussions.

\end{document}